\documentclass[fleqn]{article}

\usepackage{amsmath,amssymb}
\usepackage{graphicx}

\def\tref#1#2#3{ {#1} (#2) #3 }
\def\bref#1{(\ref{#1})}
\def\unt#1{\,{\rm #1}}
\newcommand{\nms}{\negmedspace}

\begin{document}

\title{
{\sf Instanton Effects on the Role of the
Low-Energy Theorem for the Scalar Gluonic Correlation Function} }

\author{
D.\ Harnett, T.G.\ Steele
\thanks{email: Tom.Steele@usask.ca}
\\
{\sl Department of Physics and Engineering Physics}\\
{\sl University of Saskatchewan}\\
{\sl Saskatoon, Saskatchewan S7N 5E2, Canada.}\\[10pt]
V.\ Elias
\thanks{email: velias@julian.uwo.ca}\\
{\sl Department of Applied Mathematics}\\
{\sl University of Western Ontario}\\
{\sl London, Ontario N6A 5B7, Canada.}
}
\maketitle
\begin{abstract}
Instanton contributions to the Laplace sum-rules for correlation functions of scalar gluonic currents
are calculated. The role of the constant low-energy theorem term, whose substantial contribution is 
unique to the leading Laplace sum-rule ${\cal L}_{-1}$, is shown to be diminished by instanton contributions,   significantly increasing the resulting mass bounds for the ground state of scalar 
gluonium and improving compatibility with results from higher-weight sum-rules.  
\end{abstract}

\section{Introduction}\label{intro_sec}
In the chiral limit of $n_f$ quarks, the low-energy theorem (LET) for scalar gluonic correlation functions is 
\cite{NSVZ}
\begin{equation}
\Pi(0)=\lim_{Q\rightarrow 0} \Pi\left(Q^2\right)=\frac{8\pi}{\beta_0}\left\langle J\right\rangle\quad ,
\label{let}
\end{equation}
where
\begin{eqnarray}
& &\Pi\left(Q^2\right)=i\int d^4x\, e^{iq\cdot x}\left\langle O \vert T\left[ J(x) J(0)\right] \vert O \right\rangle
\quad ,\quad Q^2=-q^2>0
\label{corr_fn}\\
& &J(x)=-\frac{\pi^2}{\alpha\beta_0}\beta\left(\alpha \right)G^a_{\mu\nu}(x)G^a_{\mu\nu}(x)
\label{current}\\
& &\beta\left(\alpha\right) =\nu^2\frac{d}{d\nu^2}\left(\frac{\alpha(\nu)}{\pi}\right)=
-\beta_0\left(\frac{\alpha}{\pi}\right)^2-\beta_1\left(\frac{\alpha}{\pi}\right)^3+\ldots
\label{beta}\\
& &\beta_0=\frac{11}{4}-\frac{1}{6} n_f\quad ,\quad
\beta_1=\frac{51}{8}-\frac{19}{24}n_f
\quad .
\label{beta_coeffs}
\end{eqnarray}
The current $J(x)$ is renormalization group (RG) invariant for massless quarks \cite{RG}, 
and its normalization has been chosen 
so that to lowest order in $\alpha$
\begin{equation}
J(x)=\alpha G^a_{\mu\nu}(x)G^a_{\mu\nu}(x)\left[1+\frac{\beta_1}{\beta_0}\frac{\alpha}{\pi}
+{\cal O}\left(\alpha^2\right)\right]
\equiv \alpha G^2(x)\left[1+\frac{\beta_1}{\beta_0}\frac{\alpha}{\pi}+{\cal O}\left(\alpha^2\right)\right]
\quad .
\label{current_exp}
\end{equation}

Most applications of dispersion relations in sum-rules are designed to remove
dependence on low-energy subtraction constants.  However, knowledge of the LET 
for gluonic correlation functions permits the possibility of sum-rules that contain
explicit dependence on the LET subtraction constant $\Pi(0)$.  
For example, the dispersion relation appropriate to the 
asymptotic (perturbative) behaviour of the correlation function \bref{corr_fn} is
\cite{dispersion}
\begin{equation}
\Pi\left(Q^2\right)=\Pi(0)+Q^2\Pi'(0)+\frac{1}{2}Q^4\Pi''(0)
-Q^6\frac{1}{\pi}\int\limits_{t_0}^\infty
dt\,\frac{ \rho(t)}{t^3\left(t+Q^2\right)} \quad .
\label{disp_rel}
\end{equation}
where $\rho(t)$ is the hadronic spectral function with physical threshold $t_0$ 
appropriate to the quantum numbers of the
current used to construct the correlation function.

Unfortunately, direct application of the dispersion relation \bref{disp_rel} is not possible because the
theoretical (perturbative) calculation of $\Pi\left(Q^2\right)$ contains a field-theoretical divergence proportional 
to $Q^4$.  A related problem is the significant contribution of excited states and the QCD continuum 
to the integral of $\rho(t)$ in \bref{disp_rel}.  Enhancement of
the lowest-lying resonance contribution in applications to light hadronic systems  
requires greater high-energy suppression of this integral.

The established technique for dealing with these issues is the Laplace sum-rules \cite{SVZ}. A family
of  Laplace sum-rules can be obtained from the dispersion relation \bref{disp_rel} through the  
Borel transform operator $\hat B$,
\begin{equation}
\hat B\equiv 
\lim_{\stackrel{N,~Q^2\rightarrow \infty}{N/Q^2\equiv \tau}}
\frac{\left(-Q^2\right)^N}{\Gamma(N)}\left(\frac{d}{dQ^2}\right)^N
\label{borel}
\end{equation}  
which has the following useful properties in the construction of the Laplace sum-rules:
\begin{eqnarray}
& &\hat B\left[a_0+a_1Q^2+\ldots a_m Q^{2m}\right]=0\quad,\quad (m~{\rm finite})
\label{borel_poly}\\
& & \hat B \left[ \frac{Q^{2n}}{t+Q^2}\right]=\tau \left(-1\right)^nt^ne^{-t\tau}  \quad,\quad n=0,~1,~2,\ldots ~ 
(n~{\rm finite})
\label{borel_exp}
\end{eqnarray}
The theoretically-determined quantity
\begin{eqnarray}
{\cal L}_k(\tau)\equiv\frac{1}{\tau}\hat B\left[\left(-1\right)^k Q^{2k}\Pi\left(Q^2\right)\right]\quad ,  
\label{laplace}
\end{eqnarray}
leads to the following family of Laplace sum-rules, after application of  $\hat B$ to the 
dispersion relation 
\bref{disp_rel} weighted by the appropriate power of $Q^2$:
\begin{eqnarray}
{\cal L}_k(\tau)&=&\frac{1}{\tau}\hat B\left[\left(-1\right)^kQ^{2k}\Pi(0)+
\left(-1\right)^kQ^{2k+2}\Pi'(0)+\left(-1\right)^k\frac{1}{2}Q^{2k+4}\Pi''(0)\right]
\nonumber\\
& &\quad-\frac{1}{\pi}\int\limits_{t_0}^\infty dt\,\frac{1}{\tau t^3}\hat B\left[\left(-1\right)^k
\frac{Q^{2k+6}}{\left(t+Q^2\right)}\right]\rho(t)
\label{lap_sr_gen}
\end{eqnarray}
There are some important constraints on $k$ that will lead to sum-rules with predictive power.  Since 
the perturbative prediction of $\Pi\left(Q^2\right)$ contains divergent constants multiplied by $Q^4$, 
the sum-rules ${\cal L}_k(\tau)$ where this contribution is absent require $k\ge -2$.  
However, the  low-energy constants $\Pi'(0)$ and $\Pi''(0)$ are {\em not} determined by the LET [{\it i.e.} only the 
quantity $\Pi(0)$ appears in \bref{let}].  Hence the sum-rules ${\cal L}_k(\tau)$ which 
will be independent of 
$\Pi'(0)$ and $\Pi''(0)$ must satisfy $k\ge -1$, and  only the $k=-1$  sum-rule will contain dependence on the 
LET-determined quantity $\Pi(0)$: 
\begin{eqnarray}
{\cal L}_{-1}(\tau)=-\Pi(0)+\frac{1}{\pi}\int\limits_{t_0}^\infty
dt\,\frac{1}{t} e^{-t\tau}\rho(t)
\label{lap_gen_m1}\\
{\cal L}_{k}(\tau)=\frac{1}{\pi}\int\limits_{t_0}^\infty
dt\,t^k e^{-t\tau}\rho(t)\quad ,\quad k>-1
\label{lap_gen_k}
\end{eqnarray}

The ``resonance(s) plus continuum''  model is used to 
represent the hadronic physics phenomenology  contained in $\rho(t)$  in 
(\ref{lap_gen_m1}--\ref{lap_gen_k}) \cite{SVZ}.  In this model,  
 hadronic physics is (locally) dual to the theoretical QCD prediction
for energies above the continuum threshold $t=s_0$:
\begin{equation}
\rho(t)\equiv\theta\left(s_0-t\right)\rho^{had}(t)+\theta\left(t-s_0\right){\rm Im}\Pi^{QCD}(t)
\label{res_plus_cont}
\end{equation} 
The contribution of the QCD continuum to the sum-rules is denoted by
\begin{equation}
c_k\left(\tau,s_0\right)=\frac{1}{\pi}\int\limits_{s_0}^\infty
dt\,t^k e^{-t\tau}{\rm Im}\Pi^{QCD}(t)\quad .
\label{continuum}
\end{equation}
Since the continuum contribution is determined by QCD, it is usually combined with the theoretical 
quantity ${\cal L}_k\left(\tau\right)$
\begin{equation}
{\cal S}_k\left(\tau,s_0\right)\equiv {\cal L}_k\left(\tau\right)-c_k\left(\tau,s_0\right)\quad ,
\label{full_sr}
\end{equation}
resulting in the following Laplace sum-rules relating QCD to hadronic physics phenomenology:
\begin{eqnarray}
{\cal S}_{-1}\left(\tau,s_0\right)=-\Pi(0)+\frac{1}{\pi}\int\limits_{t_0}^{s_0}
dt\,\frac{1}{t} e^{-t\tau}\rho^{had}(t)
\label{lap_m1}\\
{\cal S}_{k}\left(\tau,s_0\right)=\frac{1}{\pi}\int\limits_{t_0}^{s_0}
dt\,t^k e^{-t\tau}\rho^{had}(t)\quad ,\quad k>-1
\label{lap_k}
\end{eqnarray}
The property
\begin{equation}
\lim_{s_0\to\infty}c_k\left(\tau,s_0\right)=0
\label{cont_limit}
\end{equation}
implies that the   
sum-rules \bref{full_sr} and \bref{lap_gen_k} are identical in the $s_0\rightarrow\infty$ limit.
\begin{equation}
 \lim_{s_0\to\infty}{\cal S}_{k}\left(\tau,s_0\right)=
{\cal L}_k\left(\tau\right)
\label{sr_limit}
\end{equation}

The only appearance of the $\Pi(0)$ term is in the $k=-1$ sum-rule, and as first noted in
\cite{NSVZ_glue}, this LET term comprises a significant  contribution in the $k=-1$ sum-rule.
From the significance of this scale-independent term one can ascertain the  
important qualitative role of the LET in  sum-rule phenomenology.  
To see this role, we first model the hadronic contributions 
 $\rho^{had}(t)$ using the narrow resonance approximation
\begin{equation}
\frac{1}{\pi}\rho^{had}(t)=\sum_r F_r^2m_r^2\delta\left(t-m_r^2\right)\quad ,
\label{narrow}
\end{equation}
where the  sum over $r$ represents a sum over sub-continuum resonances of mass $m_r$.  The quantity
 $F_r$ is the coupling strength of the resonance to the vacuum through the gluonic current $J(0)$, so
the sum-rule for scalar gluonic currents probes  scalar gluonium states.  
In the narrow-width approximation the Laplace sum-rules  (\ref{lap_m1}--\ref{lap_k})
become
\begin{eqnarray}
& &{\cal S}_{-1}\left(\tau,s_0\right)+\Pi(0)=
\sum_r F_r^2 e^{-m_r^2\tau}
\label{lap_m1_res}\\
& &{\cal S}_{k}\left(\tau,s_0\right)=
\sum_r F_r^2m_r^{2k+2} e^{-m_r^2\tau}\quad ,\quad k>-1\quad .
\label{lap_k_res}
\end{eqnarray}
Thus if the (constant) LET term is a significant contribution on the theoretical side of \bref{lap_m1_res}, then
the left-hand side of \bref{lap_m1_res} will exhibit reduced $\tau$ dependence relative to other theoretical 
contributions.  To reproduce this diminished $\tau$ dependence, the phenomenological 
({\it i.e.} right-hand)
side must 
contain a  light resonance with a coupling larger than or comparable to the heavier resonances.  
By contrast, the absence of the $\Pi(0)$ (constant) term in $k>-1$ sum-rules leads to stronger $\tau$ dependence
which is balanced on the phenomenological side  by suppression 
of the lightest resonances 
via the additional powers of $m_r^2$  occurring in \bref{lap_k_res}.  Thus if $\Pi(0)$ is found to dominate 
${\cal S}_{-1}\left(\tau,s_0\right)$, then one would expect qualitatively different results from analysis
of the $k=-1$ and $k>-1$ sum-rules.  

Such distinct conclusions drawn from different sum-rules can be legitimate.  
In the pseudoscalar quark sector, the lowest sum-rule is dominated by the pion, and the low  mass of the pion 
is evident from the  minimal $\tau$ dependence
in the lowest sum-rule. By contrast,  the first subsequent sum-rule has an important contribution from the 
$\Pi(1300)$ \cite{pi_1300},  as the pion contribution is suppressed by its low mass, resulting in the 
significant $\tau$ dependence of the next-to-lowest sum-rule.

In the absence of instantons \cite{instanton}, 
explicit sum-rule analyses of scalar gluonium \cite{dispersion,NSVZ_glue,glue} uphold the above 
generalization---those which include the
$k=-1$ sum-rule find a light (less than or on the order of the $\rho$ mass) gluonium state, 
and those which omit the $k=-1$  sum-rule find a state with a mass greater than $1\unt{GeV}$. 
The prediction of a light gluonium state  would have interesting 
phenomenological consequences as a state which could be identified with the $f_0(400-1200)/\sigma$ 
meson \cite{ochs}.   However, a detailed treatment of instanton contributions is essential in assessing
the viability of such a two-resonance-scale scenario (as evident in the sum-rule analysis of the $q\bar q$ 
pseudoscalar channel \cite{pi_1300}) in the scalar gluonium channel.

Seminal work by Shuryak \cite{shuryak} in the instanton liquid model \cite{ins_liquid} has indicated how
an asymptotic ($\rho_c/\sqrt{\tau}\gg 1$) expression for the instanton contribution to the $k=-1$ sum-rule
may serve to compensate for that sum-rule's LET component and bring the predicted scalar gluonium mass in
line with subsequent lattice estimates ($\sim 1.6\unt{GeV}$ \cite{lattice}).  Recent work by Forkel \cite{forkel} has addressed
in detail instanton effects on scalar gluonium mass predictions from higher-weight sum-rules and has also corroborated
lattice estimates.  However, the overall consistency of the $k=-1$ sum-rule, which is sensitive to the low-energy
theorem term, and $k\ge 0$ sum-rules, which are not, has not been addressed quantitatively.

In Section~\ref{lap_sec}, we explicitly calculate the instanton contributions to 
Laplace sum-rules of scalar gluonic currents.
We pay particular attention to the $k=-1$ sum-rule and demonstrate that
instanton contributions partially cancel against the LET constant $\Pi(0)$ and serve
to appreciably diminish its dominance of this leading order sum-rule.
The phenomenological implications of this partial cancellation
are investigated in Section~\ref{phenom_sec}, and a discussion
relating our work to other analyses of instanton effects in the scalar gluonium channel is presented
in Section~\ref{discuss_sec}.

\section{Instanton Effects in the Laplace Sum-Rules}
\label{lap_sec}  
The field-theoretical (QCD) calculation of $\Pi\left(Q^2\right)$ consists of perturbative (logarithmic) corrections 
known to three-loop order ($\overline{{\rm MS}}$ scheme) in the chiral limit of $n_f=3$ massless quarks 
\cite{glue_pt}, 
QCD vacuum effects of infinite correlation length 
parameterized by  the power-law contributions from the QCD vacuum condensates 
\cite{NSVZ_glue,condensate2}, 
\footnote{The calculation of one-loop contributions proportional to $\left\langle J\right\rangle$ 
in \protect\cite{condensate2} have been extended 
non-trivially
to $n_f=3$ from $n_f=0$, and the operator basis
has been changed from $\left\langle \alpha G^2\right\rangle$ to $\left\langle J\right\rangle$. 
}
and QCD vacuum effects of finite correlation length devolving from instantons \cite{instanton_corr}
\begin{equation}
\Pi\left(Q^2\right)=
\Pi^{pert}\left(Q^2\right)+\Pi^{cond}\left(Q^2\right)+\Pi^{inst}\left(Q^2\right)\quad ,\quad 
\label{QCD_corr_fn}
\end{equation}
with
\begin{enumerate}
\item $\ldots$ the perturbative contribution (ignoring divergent terms proportional to $Q^4$) given by
\begin{eqnarray}
& &\Pi^{pert}\left(Q^2\right)=Q^4\log\left(\frac{Q^2}{\nu^2}\right)\left[
a_0+a_1\log\left(\frac{Q^2}{\nu^2}\right)+a_2\log^2\left(\frac{Q^2}{\nu^2}\right)
\right]
\label{pi_pert}\\
& &a_0=-2\left(\frac{\alpha}{\pi}\right)^2\left[1+\frac{659}{36}\frac{\alpha}{\pi}+
247.480\left( \frac{\alpha}{\pi}\right)^2\right]
~ , ~  
a_1=2\left(\frac{\alpha}{\pi}\right)^3\left[ \frac{9}{4}+65.781\frac{\alpha}{\pi}\right]
~ , ~ 
a_2=-10.1250\left(\frac{\alpha}{\pi}\right)^4
\nonumber
\end{eqnarray}

\item $\ldots$ the condensate contributions given by 
\begin{eqnarray}
& &\Pi^{cond}\left(Q^2\right)=
\left[ b_0+b_1\log\left(\frac{Q^2}{\nu^2}\right)\right]\left\langle  J\right\rangle
+c_0\frac{1}{Q^2}\left\langle {\cal O}_6\right\rangle+d_0\frac{1}{Q^4}\left\langle {\cal O}_8\right\rangle
\label{pi_cond}\\
& &b_0=4\pi\frac{\alpha}{\pi}\left[ 1+ \frac{175}{36}\frac{\alpha}{\pi}\right]
\quad ,\quad b_1=-9\pi\left(\frac{\alpha}{\pi}\right)^2
\quad ,\quad
c_0=8\pi^2\left(\frac{\alpha}{\pi}\right)^2\quad ,\quad d_0=8\pi^2\frac{\alpha}{\pi}
\label{GG_coeffs}\\
& &\left\langle {\cal O}_6\right\rangle=
\left\langle g f_{abc}G^a_{\mu\nu}G^b_{\nu\rho}G^c_{\rho\mu}\right\rangle
\quad ,\quad
\left\langle {\cal O}_8\right\rangle=14\left\langle\left(\alpha f_{abc}G^a_{\mu\rho}G^b_{\nu\rho}\right)^2\right\rangle
-\left\langle\left(\alpha f_{abc}G^a_{\mu\nu}G^b_{\rho\lambda}\right)^2\right\rangle
\label{higher_dim_conds}
\end{eqnarray}

\item $\ldots$ and the instanton contribution given by 
\begin{equation}
\Pi^{inst}\left(Q^2\right)=
32\pi^2Q^4\int \rho^4 \left[K_2\left(\rho\sqrt{Q^2}\right)\right]^2  dn(\rho)\quad ,
\label{pi_inst}
\end{equation}
where $K_2(x)$ represents a modified Bessel function \cite{abramowitz}.
\end{enumerate}
The strong coupling constant $\alpha$ is understood to be the running coupling at the renormalization 
scale $\nu$, 
and renormalization group improvement of the Laplace sum-rules implies that $\nu^2=1/\tau$ \cite{RG_improve}. 
The instanton contributions 
represent a calculation with non-interacting instantons 
of size $\rho$, with subsequent integration over the instanton density distribution $n(\rho)$. 
\footnote{A factor of 2 to include the sum of instanton and anti-instanton contributions has  been 
included in (\protect\ref{QCD_corr_fn}).} 
The theoretical contributions to the Laplace sum-rules corresponding to \bref{QCD_corr_fn} are
\begin{equation}
{\cal L}_k(\tau)=
{\cal L}_k^{pert}(\tau)+{\cal L}_k^{cond}(\tau)+{\cal L}_k^{inst}(\tau)\quad . 
\label{laplace_ports}
\end{equation}

An alternative to the
direct calculation of the Laplace sum-rules through the definition of $\hat B$  in \bref{borel} is obtained through 
an identity relating the Borel and Laplace transform \cite{borel_ref}
\begin{eqnarray}
& &f\left(Q^2\right)=\int\limits_0^\infty d\tau F(\tau) e^{-Q^2\tau}\equiv{\cal L}\left[ F(\tau)\right]
\quad \Longrightarrow\quad \frac{1}{\tau}\hat B\left[ f\left(Q^2\right)\right]
=F(\tau)={\cal L}^{-1}
\left[ f\left(Q^2\right)\right]
\label{borel_laplace}\\
& &{\cal L}^{-1}
\left[ f\left(Q^2\right)\right]=\frac{1}{2\pi i}\int\limits_{b-i\infty}^{b+i\infty}
f\left(Q^2\right) e^{Q^2\tau}\,dQ^2
\label{inv_lap_def}
\end{eqnarray}
where the real parameter $b$ in the definition \bref{inv_lap_def} of the inverse Laplace 
transform must be chosen so that $f\left(Q^2\right)$ is analytic to the right of the 
contour of integration in the complex plane.
Using the result \bref{borel_laplace}, the Laplace sum-rules \bref{laplace} can be  
obtained from an inverse Laplace transform of 
the theoretically-determined correlation function:
\begin{equation}
{\cal L}_k(\tau)={\cal L}^{-1}\left[
\left(-1\right)^k Q^{2k}\Pi\left(Q^2\right)
\right]\quad .
\label{inv_lap}
\end{equation}

\begin{figure}[htb]
\centering
\includegraphics[scale=0.5,angle=270]{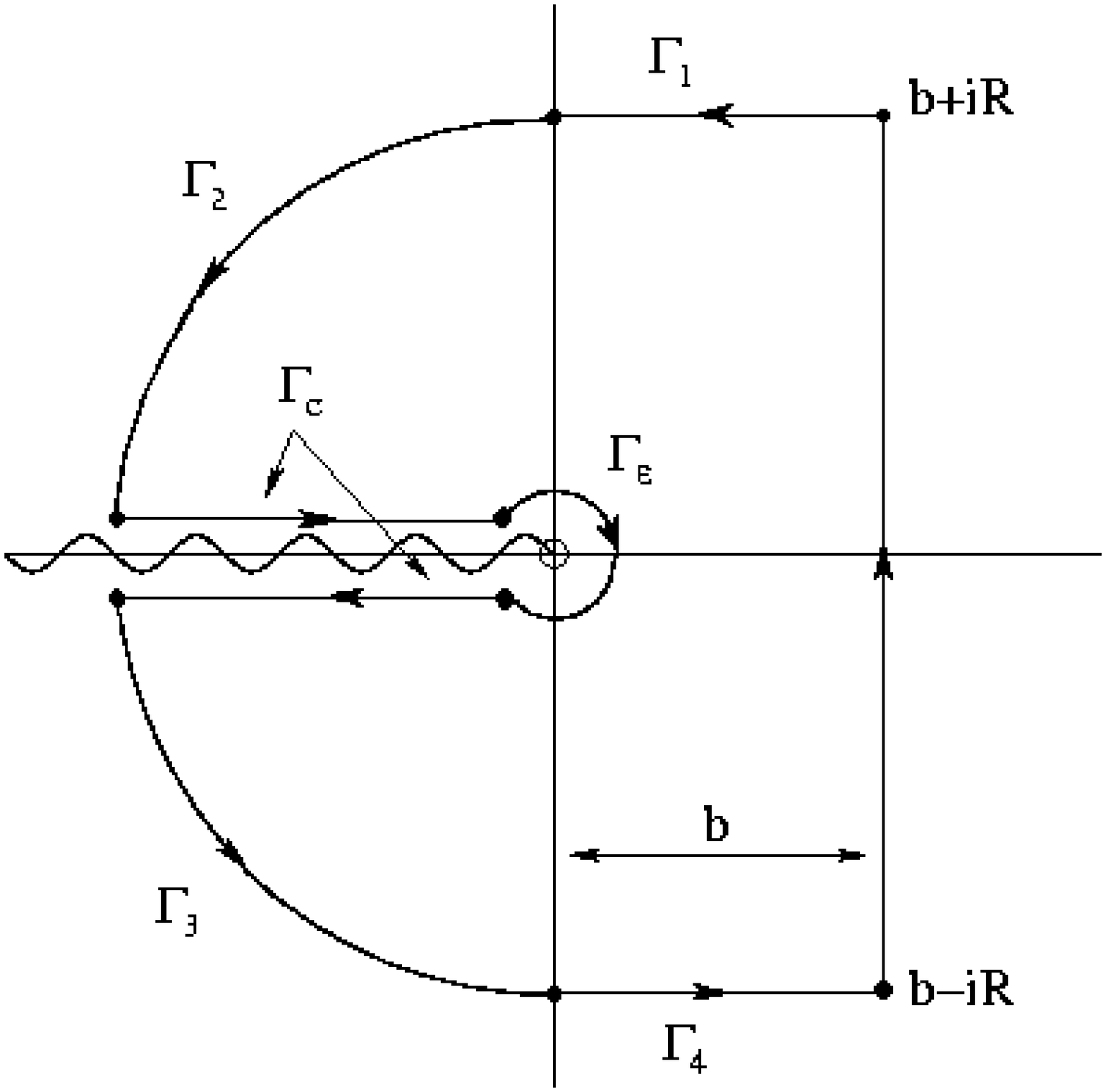}
\caption{Closed contour $C(R)$ used to 
obtain the inverse Laplace transform (\protect\ref{inv_lap}) defining the Laplace sum-rules. 
The  inner circular segment $\Gamma_\epsilon$ has a radius of $\epsilon$, and the outer circular segments 
$\Gamma_2$ and $\Gamma_3$
have a radius $R$.
The wavy line 
on the negative real axis denotes the branch cut of $\Pi(z)$, 
and the linear segments of the contour above and below the branch cut are denoted by $\Gamma_c$.
}
\label{cont_fig1}
\end{figure}

In the complex $Q^2$ plane where the inverse Laplace transform \bref{inv_lap_def} is calculated, 
the QCD expression \bref{QCD_corr_fn} for the correlation function 
$\Pi\left(Q^2\right)$ 
is analytic apart from a branch point at $Q^2=0$ with a branch cut 
extending to infinity along the negative-real-$Q^2$ axis. 
Consequently, analyticity to the right of the contour in \bref{inv_lap_def} implies that
 $b>0$.
Consider the contour
$C(R)$ in Figure \ref{cont_fig1}; $\Pi\left(Q^2\right)$ is analytic within and on $C(R)$ and so with $z=Q^2$
\begin{equation}
0=
\frac{1}{2\pi i}\oint\limits_{C(R)} \left(-z\right)^k e^{z\tau}\Pi(z) dz\quad ,
\label{cont_result}
\end{equation}
which leads to 
\begin{equation}
\frac{1}{2\pi i}\int\limits_{b-iR}^{b+iR} \left(-z\right)^k e^{z\tau}\Pi(z) dz
=-\frac{1}{2\pi i}\int\limits_{\Gamma_1+\ldots +\Gamma_4}\!\!\! \left(-z\right)^k e^{z\tau}\Pi(z) dz
-\frac{1}{2\pi i}\int\limits_{\Gamma_c+\Gamma_\epsilon}\!\! \left(-z\right)^k e^{z\tau}\Pi(z) dz
\quad .
\label{borel_inv_1}
\end{equation}  
Taking the limit as $R\to\infty$, which requires use of the asymptotic behaviour  of the modified Bessel 
function \cite{abramowitz}
\begin{equation}
K_2\left(z\right)\sim\sqrt{\frac{\pi}{2z}}e^{-z}\quad ;\quad |z|\gg 1~,~\left|{\rm arg}(z)\right|\le\frac{\pi}{2}
\end{equation}
the individual integrals over $\Gamma_{1\ldots 4}$ are found to 
vanish, resulting in the following expression for the Laplace sum-rule.
\begin{equation}
{\cal L}_{k}\left(\tau\right)=
\frac{1}{2\pi i}\int\limits_\epsilon^\infty t^ke^{-t\tau}
\left[\Pi\left(te^{-i\pi}\right)-\Pi\left(te^{i\pi}\right)\right]  dt
+\frac{1}{2\pi }\int\limits_{-\pi}^\pi \left(-1\right)^k \exp{\left(\epsilon e^{i\theta}\tau\right)}
\epsilon^{k+1}e^{i (k+1)\theta}\Pi\left(\epsilon e^{i\theta}\right)   d\theta
\label{simpl_inv_borel}
\end{equation}

Perturbative and QCD condensate contributions to the Laplace sum-rules are well known 
\cite{dispersion,NSVZ_glue,glue}, and serve as a consistency check for the conventions used to determine the 
instanton contribution
through \bref{simpl_inv_borel}.  Keeping in mind the $k\ge -1$ constraint established previously, we see 
that  the perturbative contributions to the $\theta$ integral in \bref{simpl_inv_borel}
are zero in the limit as $\epsilon\to 0$, leaving only the anticipated integral of the discontinuity 
across the branch cut [{\it i.e.} ${\rm Im}\Pi^{pert}(t)$] to determine the following perturbative contributions 
to the Laplace sum-rule.
\begin{equation}
{\cal L}^{pert}_{k}\left(\tau\right)=\int\limits_0^{\infty}
t^{k+2}e^{-t\tau} \left[-a_0-2a_1\log\left(\frac{t}{\nu^2}\right)
+a_2\left(\pi^2-3\log^2\left(\frac{t}{\nu^2}\right)\right)\right]\,dt
\label{sk_pert}
\end{equation}

The QCD condensate terms  proportional to $b_0$, $c_0$ and $d_0$ in the correlation function 
$\Pi(z)$ do not have a branch discontinuity, so their contribution to the Laplace sum-rule
arises solely from the contour $\Gamma_\epsilon$ (represented by the term in 
\bref{simpl_inv_borel} with 
the $\theta$ integral), 
and can be evaluated using the result
\begin{equation}
-\frac{1}{2\pi i}\int\limits_{\Gamma_\epsilon}\frac{e^{z\tau}}{z^n}dz=
\left\{
\begin{array}{l}
0\quad ,\quad n=0,~-1,~-2,\ldots\\
\frac{\tau^{n-1}}{(n-1)!}\quad,\quad n=1,~2,~3,\ldots 
\end{array}
\right.
\label{power_law_sk}
\end{equation}     
The QCD condensate term proportional to $b_1$ requires a more careful treatment. If $\Pi(z)$
is replaced with $\log\left(z/\nu^2\right)$ in \bref{simpl_inv_borel}  then we find
\begin{equation}
\frac{1}{2\pi i}\int\limits_{b-i\infty}^{b+i\infty}\!\!\! \left(-z\right)^k e^{z\tau}
\log\left(\frac{z}{\nu^2}\right) dz
=-\int\limits_\epsilon^{\infty} t^ke^{-t\tau}
 dt
+\frac{1}{2\pi }\int\limits_{-\pi}^\pi \!\!\left(-1\right)^k \exp{\left(\epsilon e^{i\theta}\tau\right)}
\epsilon^{k+1}e^{i (k+1)\theta}\left(\log\left(\frac{\epsilon}{\nu^2}\right) +i\theta\right)   d\theta
\label{def_cont_int2}
\end{equation}
The last term in this equation will be zero in the $\epsilon\to 0$ limit except  when $k=-1$. Similarly, 
the $t$ integral is well defined in the  $\epsilon\to 0$ limit except when $k=-1$.  With $\nu^2=1/\tau$, 
and with evaluation of the $\epsilon\to 0$ limit [which, for $k=-1$, involves cancellation between the two 
integrals in \bref{def_cont_int2}] we find
\begin{equation}
\frac{1}{2\pi i}\int\limits_{b-i\infty}^{b+i\infty} \left(-z\right)^k e^{z\tau}
\log\left(\frac{z}{\nu^2}\right) dz
=\left\{\begin{array}{l}
-\int\limits_0^{\infty} t^ke^{-t\tau}dt \quad ,\quad k>-1\\
\gamma_{_E} \quad,\quad k=-1
\end{array}
\right.
\label{b1_int}
\end{equation}
where  $\gamma_{_E}\approx 0.5772$ is Euler's constant.
It is easily verified that equations \bref{b1_int}, \bref{power_law_sk}, and \bref{sk_pert} lead to the
known results \cite{dispersion,NSVZ_glue,glue} for the non-instanton contributions to the  Laplace sum-rules for scalar gluonic
currents.

To evaluate the instanton contributions to the Laplace sum-rule, we must calculate the following integral:
\begin{eqnarray}
\frac{1}{2\pi i}\int\limits_{b-i\infty}^{b+i\infty} \left(-z\right)^k e^{z\tau}z^2
\left[K_2\left(\rho\sqrt{z}\right)\right]^2 dz
&=&\frac{1}{2\pi i}\int\limits_\epsilon^{\infty} t^{k+2}e^{-t\tau}
\left[
\left[K_2\left(\rho\sqrt{t}e^{-i\pi/2}\right)\right]^2
-
\left[K_2\left(\rho\sqrt{t}e^{i\pi/2}\right)\right]^2
\right]dt
\nonumber\\
& &
+\frac{1}{2\pi }\int\limits_{-\pi}^\pi \left(-1\right)^k \exp{\left(\epsilon e^{i\theta}\tau\right)}
\epsilon^{k+3}e^{i (k+3)\theta}\left[K_2\left(\rho\sqrt{\epsilon} e^{i\theta/2}\right)\right]^2   d\theta
\label{inst_int2}
\end{eqnarray}
Simplification of \bref{inst_int2} requires the following properties of the modified Bessel
function $K_2(z)$ \cite{abramowitz}
\begin{eqnarray}
& &K_2\left(z\right)\sim\frac{2}{z^2}\quad ,\quad z\to 0
\label{asymp_K2}\\
& &K_2\left(z\right)=\left\{
\begin{array}{l}
-i\frac{\pi}{2}H_2^{(1)}\left(ze^{i\pi/2}\right)\quad ,\quad -\pi<\arg(z)\le \frac{\pi}{2}\\[5pt]
i\frac{\pi}{2}H_2^{(2)}\left(ze^{-i\pi/2}\right)\quad ,\quad -\frac{\pi}{2}<\arg(z)\le \pi
\end{array}
\right.
\label{anal_con_K2}
\end{eqnarray}
where $H_2^{(1)}(z)=J_2(z)+iY_2(z)$ and $H_2^{(2)}(z)=J_2(z)-iY_2(z)$.
The asymptotic behaviour \bref{asymp_K2} implies that the $\theta$ integral of \bref{inst_int2}
will be zero in the $\epsilon\to 0$ limit for $k>-1$ and the identity \bref{anal_con_K2}
allows evaluation of the discontinuity in the $t$ integral of~\bref{inst_int2}, leading to the
following instanton contribution to the Laplace sum-rules: 
\begin{align}
  \mathcal{L}_{-1}^{inst}\left(\tau\right) &=
  -16\pi^3\int\nms\mathrm{d}n(\rho) \rho^4\int\limits_0^\infty tJ_2\left(\rho\sqrt{t}\right)
  Y_2\left(\rho\sqrt{t}\right)\mathrm{e}^{-t\tau}\,\mathrm{d}t -128\pi^2\int\mathrm{d}n(\rho)
  \label{sm1_inst}\\
  &= -64\pi^2 \int\nms\mathrm{d}n(\rho)\ a\mathrm{e}^{-a} \left[\;(1+a)aK_0(a)
      +(2+2a+a^2)K_1(a)\;\right]\nonumber\\
  {\cal L}_k^{inst}\left(\tau\right)    &=
  -16\pi^3\int\nms\mathrm{d}n(\rho) \rho^4\int\limits_0^\infty t^{k+2}J_2\left(\rho\sqrt{t}\right)
  Y_2\left(\rho\sqrt{t}\right)\mathrm{e}^{-t\tau} \,\mathrm{d}t \quad ,\quad k>-1
  \label{sk_inst}\\
  &= \left\{
     \begin{aligned}
       \; & 128\pi^2\int\nms\mathrm{d}n(\rho)\ \frac{a^4\mathrm{e}^{-a}}{\rho^2}
       \left[\; 2aK_0(a)+(1+2a)K_1(a)\;\right]\quad,\quad k=0  \\
       \; & 256\pi^2 \int\nms\mathrm{d}n(\rho)\ \frac{a^5\mathrm{e}^{-a}}{\rho^4}
       \left[\; (9-4a)aK_0(a) + (3+7a-4a^2)K_1(a)\;\right]\quad,\quad k=1
     \end{aligned}
     \right. \nonumber
\end{align}
where $a\equiv\rho^2/(2\tau)$ and where $K_n$ is the modified Bessel function of the
first kind of order $n$ (\textit{c.f.}~\cite{abramowitz}). 
Observe the symmetry between~(\ref{sm1_inst}) and~(\ref{sk_inst}) broken  by the
term $-128\pi^2\int\mathrm{d}n({\rho})$ appearing in~(\ref{sm1_inst})---a
term which corresponds to the second integral on the right-hand side of~(\ref{inst_int2})
and which is nonzero \emph{only} for $k=-1$. 
Conversely, we note that naively substituting $k=-1$ into~(\ref{sk_inst}) leads to
an incorrect expression for the instanton contribution to the leading order sum-rule.  
This asymmetric role  of  the instanton contributions to 
$k=-1$ and $k>-1$ sum-rules is also a property of the LET as illustrated 
in~(\ref{lap_gen_m1},\ref{lap_gen_k}).

As discussed in Section~\ref{intro_sec}, we wish to determine whether the leading
order sum-rule $\mathcal{L}_{-1}$ might support the existence of a lowest-lying
resonance whose presence is mass-suppressed in subsequent higher order $k>-1$ sum-rules.
Such is indeed the case, for example, for the pion within sum-rules based on a
pseudoscalar $\overline q\gamma_5q$ current. Correspondingly, one might anticipate
the identification of the lowest-lying scalar gluonium state with a
$500\mbox{--}600\,{\rm MeV}$ $\sigma$ ({\it i.e.} the lower-mass range of the $f_0(400-1200)$) 
resonance~\cite{ochs}
whose contribution to higher order sum-rules is suppressed by additional
factors of $m^2_{\sigma}$ [\textit{i.e.}\ the additional factors of $m^2_r$
in~(\ref{lap_k_res})], a
scenario analogous to the $m^2_{\pi}$  suppression of pion contributions to the
pseudoscalar $\overline q\gamma_5q$ sum-rules~\cite{pi_1300}. As
already noted in Section~\ref{intro_sec}, the LET constant in the absence of instantons
supports this scenario for a sub-GeV scalar glueball.

\begin{figure}[htb]
  \centering
  \includegraphics[scale=0.7]{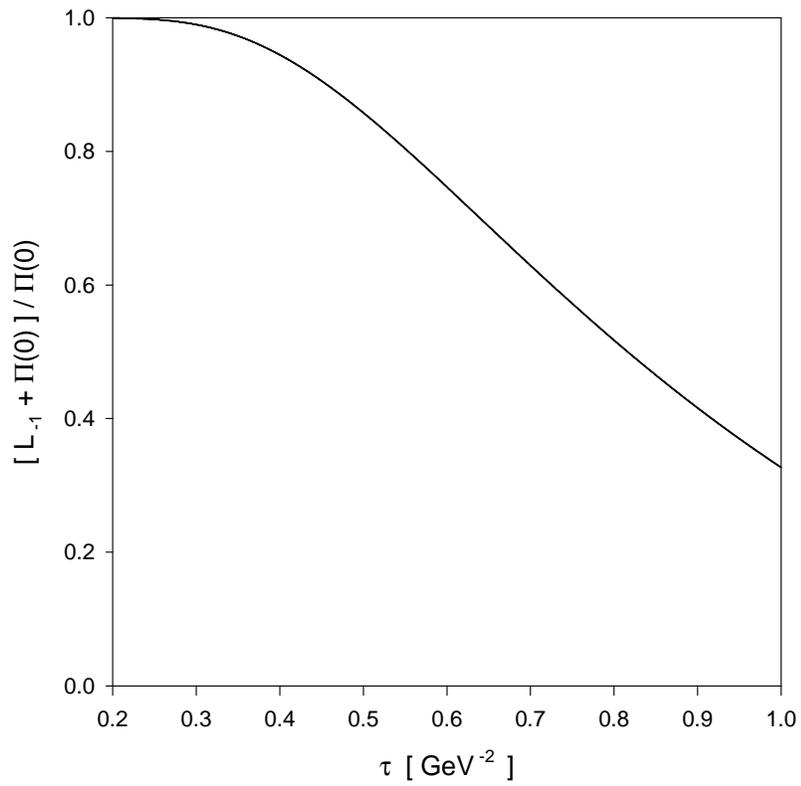}
  \caption{\label{mitigation} 
The quantity $\left[\Pi(0)+\mathcal{L}^{inst}_{-1}(\tau)\right]/\Pi(0)$ is plotted as a function of $\tau$,
           illustrating that  instanton effects
             cancel a significant portion of the LET term in the $k=-1$ sum-rule.   }
\end{figure}

     However, the instanton contribution to the $k = -1$ sum-rule is opposite in sign and
comparable in magnitude to the LET subtraction constant $\Pi(0)$, thereby ameliorating this term's
dominance of the lowest order sum-rule. For example, the contribution of instanton and LET
terms to the $k = -1$ sum rule in the dilute instanton liquid (DIL) model~\cite{ins_liquid},
\begin{equation}
dn(\rho)=n_c\delta\left(\rho-\rho_c\right)d\rho\quad ;\quad n_c=8\times 10^{-4}\unt{GeV^4}\quad ,\quad
\rho_c=\frac{1}{600\unt{MeV}}
\label{ins_liquid_params}
\end{equation}
 renders trivial the remaining integrations in~(\ref{sm1_inst}--\ref{sk_inst}).
If we approximate $\langle J\rangle$ by
$\left\langle \alpha G^2\right\rangle$
and employ a recently determined value of the gluon condensate \cite{GG_ref}
\begin{equation}
\left\langle \alpha G^2\right\rangle=(0.07\pm 0.01) \unt{GeV^4}
\label{GG_cond}
\end{equation}
we obtain via~(\ref{let}) and~(\ref{beta_coeffs}) an $n_f=3$ estimate of  the
LET subtraction constant:
\begin{equation}
\label{PI_approx}
  \Pi(0)\approx\frac{32\pi}{9}\left\langle \alpha G^2\right\rangle
        \approx (0.78\pm0.11)\ \unt{GeV^4}\quad.
\end{equation}
In Figure~\ref{mitigation}, we use~(\ref{sm1_inst}) and the central value of~(\ref{PI_approx}) [with $n_c$ and $\rho_c$
given in~(\ref{ins_liquid_params})] to plot
\begin{equation}
   \frac{\Pi(0) + \mathcal{L}^{inst}_{-1}(\tau)}{\Pi(0)}
\end{equation}
as a function $\tau$. We note that, as anticipated, instanton effects do indeed
significantly reduce the impact of the LET on the $k=-1$ sum-rule:
anywhere from 20--65\% for $\tau$
ranging between $0.6\ \mathrm{GeV}^{-2}$ and  $1.0\ \mathrm{GeV}^{-2}$.
Recalling that the dominance of the LET
over ${\cal S}_{-1}$ is responsible for the discrepancy in gluonium mass scales in the analysis of the
$k=-1$ and $k>-1$ sum-rules, we see that suppression of the LET by instanton effects
could reconcile this discrepancy, a possibility which is investigated further in the next section.

\section{Phenomenological Impact of Instanton Effects in the Laplace Sum-Rules}
\label{phenom_sec}
Ratios of Laplace sum-rules provide a simple technique for extracting the mass of the lightest
(narrow) resonance probed by the sum-rules. If only the lightest resonance  (of mass $m$) is included in  
\bref{lap_m1_res} and \bref{lap_k_res}, then for the first few sum-rules we see that
\begin{eqnarray}
& &\frac{{\cal S}_1\left(\tau,s_0\right)}{{\cal S}_0\left(\tau,s_0\right)}=m^2
\label{r1_r0_ratio}\\
& &\frac{{\cal S}_0\left(\tau,s_0\right)}{{\cal S}_{-1}\left(\tau,s_0\right)+\Pi(0)}=m^2\quad .
\label{r0_rm1_ratio}
\end{eqnarray}
This method of predicting the mass $m$ requires optimization of $s_0$ to minimize the  $\tau$ dependence that can occur in the 
sum-rule ratios.  However, a qualitative analysis which avoids these optimization issues
occurs in the $s_0\to\infty$ limit where bounds on the mass $m$ can also be obtained.
These bounds originate from inequalities satisfied on the hadronic physics side of the 
sum-rule because of the positivity of $\rho^{had}(t)$.  For example,
\begin{eqnarray}
& &\frac{1}{\pi}\int\limits_{t_0}^{s_0}
dt\,t e^{-t\tau}\rho^{had}(t)=\frac{1}{\pi}\int\limits_{t_0}^{s_0}
dt\,\left(t-s_0+s_0\right) e^{-t\tau}\rho^{had}(t)\le s_0
\frac{1}{\pi}\int\limits_{t_0}^{s_0}
dt\, e^{-t\tau}\rho^{had}(t)\nonumber\\
& &\Longrightarrow
{\cal S}_1\left(\tau,s_0\right)\le s_0{\cal S}_0\left(\tau,s_0\right)\quad .
\label{s1_ineq}
\end{eqnarray}
Furthermore,  positivity of ${\rm Im}\Pi^{QCD}(t)$ leads to an inequality for the
continuum.
\begin{eqnarray}
& &\frac{1}{\pi}\int\limits_{s_0}^{\infty}
dt\,t e^{-t\tau}{\rm Im}\Pi^{QCD}(t)=\frac{1}{\pi}\int\limits_{s_0}^{\infty}
dt\,\left(t-s_0+s_0\right) e^{-t\tau}{\rm Im}\Pi^{QCD}(t)\ge s_0
\frac{1}{\pi}\int\limits_{s_0}^{\infty}
dt\, e^{-t\tau}{\rm Im}\Pi^{QCD}(t)\nonumber\\
& &\Longrightarrow
c_1\left(\tau,s_0\right)\ge s_0 c_0\left(\tau,s_0\right)
\label{c1_ineq}
\end{eqnarray}  
These inequalities can be extended to include the $k=-1$ sum-rules and continuum.
\begin{eqnarray}
& &{\cal S}_0\left(\tau,s_0\right)\le s_0\left[{\cal S}_{-1}\left(\tau,s_0\right)+\Pi(0)\right]
\label{s0_ineq}\\
& &c_0\left(\tau,s_0\right)\ge s_0 c_{-1}\left(\tau,s_0\right)
\label{c0_ineq}
\end{eqnarray}
We then see that
\begin{equation}
\frac{{\cal L}_1\left(\tau\right)}{{\cal L}_0\left(\tau\right)}
=\frac{{\cal S}_1\left(\tau,s_0\right)+c_1\left(\tau,s_0\right)}{{\cal S}_0\left(\tau,s_0\right)
+c_0\left(\tau,s_0\right)}
= \frac{{\cal S}_1\left(\tau,s_0\right)}{{\cal S}_0\left(\tau,s_0\right)}
\left[
\frac{1+\frac{c_1\left(\tau,s_0\right)}{{\cal S}_1\left(\tau,s_0\right)}}{1+\frac{c_0\left(\tau,s_0\right)}{{\cal S}_0\left(\tau,s_0\right)}}
\right]
\ge \frac{{\cal S}_1\left(\tau,s_0\right)}{{\cal S}_0\left(\tau,s_0\right)} =m^2
\label{s1_bounds}
\end{equation}
where the final inequality of \bref{s1_bounds} follows from
$c_1/{\cal S}_1\ge s_0 c_0/{\cal S}_1\ge c_0/{\cal S}_0$ via
\bref{c1_ineq} and \bref{s1_ineq}.  Similarly, we find from \bref{s0_ineq} and \bref{c0_ineq} that
\begin{equation}
\frac{{\cal L}_0\left(\tau\right)}{{\cal L}_{-1}\left(\tau\right)+\Pi(0)}
=\frac{{\cal S}_0\left(\tau,s_0\right)+c_0\left(\tau,s_0\right)}{{\cal S}_{-1}\left(\tau,s_0\right)+\Pi(0)+c_{-1}\left(\tau,s_0\right)}
\ge \frac{{\cal S}_0\left(\tau,s_0\right)}{{\cal S}_{-1}\left(\tau,s_0\right)+\Pi(0)} =m^2\quad.
\label{s0_bounds}
\end{equation}
Thus the ratios of the $s_0\to\infty$ limit of the sum-rules provide bounds on the mass in this single narrow 
resonance approximation.  Extending the
analysis to many narrow resonances alters (\ref{r1_r0_ratio}--\ref{r0_rm1_ratio}) so that the sum-rule ratios are an upper 
 bound on 
the lightest resonance, upholding the bounds  (\ref{s1_bounds}--\ref{s0_bounds}) on the 
mass $m^2$ of the lightest resonance.
\begin{eqnarray}
& &\frac{{\cal L}_1\left(\tau\right)}{{\cal L}_0\left(\tau\right)}
\ge m^2
\label{L1_bounds}\\
& &\frac{{\cal L}_0\left(\tau\right)}{{\cal L}_{-1}\left(\tau\right)+\Pi(0)}
\ge m^2
\label{L0_bounds}
\end{eqnarray}

The sum-rule bounds in (\ref{L1_bounds}--\ref{L0_bounds}) can now be employed to determine the phenomenological 
impact of the instanton contributions on the sum-rule estimates of the lightest gluonium state, and to assess whether
the suppression of the LET contribution by the instanton effects is sufficient to reduce the discrepancy between 
sum-rule analyses containing or omitting the $k=-1$ sum-rule.  
Collecting results  from equations 
(\ref{pi_cond},\ref{simpl_inv_borel}--\ref{power_law_sk},\ref{sm1_inst},\ref{sk_inst}) 
the first few 
sum-rules ${\cal L}_k(\tau)$ are
\begin{align}
\mathcal{L}_{-1}(\tau) &= \frac{1}{\tau^2}\left[-a_0+a_1\left(-2+2\gamma_E\right)
  +a_2\left(\frac{\pi^2}{2}+6\gamma_E-3\gamma_E^2\right)\right]
  +\left(-b_0+b_1\gamma_E\right)\left\langle J\right\rangle
  -c_0\tau\left\langle {\cal O}_6\right\rangle
  -d_0\frac{\tau^2}{2}\left\langle{\cal O}_8\right\rangle
  \nonumber\\
  & \hspace{3em} -64\pi^2 \int\nms\mathrm{d}n(\rho)\ a\mathrm{e}^{-a} \left[\;(1+a)aK_0(a)
      +(2+2a+a^2)K_1(a)\;\right]
\label{Lm1}\\
\mathcal{L}_{0}(\tau) &= \frac{1}{\tau^3}\left[-2a_0+a_1\left(-6+4\gamma_E\right)
  +a_2\left(\pi^2-6+18\gamma_E-6\gamma_E^2\right)\right]
  -\frac{b_1}{\tau}\left\langle J\right\rangle
  +c_0\left\langle {\cal O}_6\right\rangle
  +d_0\tau\left\langle{\cal O}_8\right\rangle
  \nonumber\\
  & \hspace{3em} +128\pi^2\int\nms\mathrm{d}n(\rho)\ \frac{a^4\mathrm{e}^{-a}}{\rho^2}
       \left[\; 2aK_0(a)+(1+2a)K_1(a)\;\right]
\label{L0}\\
\mathcal{L}_{1}(\tau) &= \frac{1}{\tau^4}\left[-6a_0+a_1\left(-22+12\gamma_E\right)
  +a_2\left(3\pi^2-36+66\gamma_E-18\gamma_E^2\right)\right]
  -\frac{b_1}{\tau^2}\left\langle J\right\rangle
  -d_0\left\langle{\cal O}_8\right\rangle
\nonumber\\
  & \hspace{3em} +256\pi^2 \int\nms\mathrm{d}n(\rho)\ \frac{a^5\mathrm{e}^{-a}}{\rho^4}
       \left[\; (9-4a)aK_0(a) + (3+7a-4a^2)K_1(a)\;\right] \quad.
\label{L1}
\end{align}
Renormalization-group
improvement has been achieved by setting $\nu^2=1/\tau$ in the correlation
function and in the (three-loop, $n_f=3$, $\overline{\rm MS}$) running coupling $\alpha$:  
\begin{eqnarray}
\frac{\alpha_s(\nu)}{\pi}&=& \frac{1}{\beta_0 L}-\frac{\bar\beta_1\log L}{\left(\beta_0L\right)^2}+
\frac{1}{\left(\beta_0 L\right)^3}\left[
\bar\beta_1^2\left(\log^2 L-\log L -1\right) +\bar\beta_2\right]
\label{alpha_hl}\\
& &L=\log\left(\frac{\nu^2}{\Lambda^2}\right)\quad ,\quad \bar\beta_i=\frac{\beta_i}{\beta_0}
\quad ,
\quad
\beta_0=\frac{9}{4}\quad ,\quad \beta_1=4\quad ,\quad \beta_2=\frac{3863}{384}
\end{eqnarray}
with 
 $\Lambda_{\overline{MS}}\approx 300\,{\rm MeV}$ for three active flavours,
consistent with current estimates of $\alpha_s(M_\tau)$ \cite{PDG,pade_tau} and matching conditions through the charm threshold 
\cite{ChetyrkinKniehl}.

\begin{figure}[htb]
\centering
\includegraphics[scale=0.7]{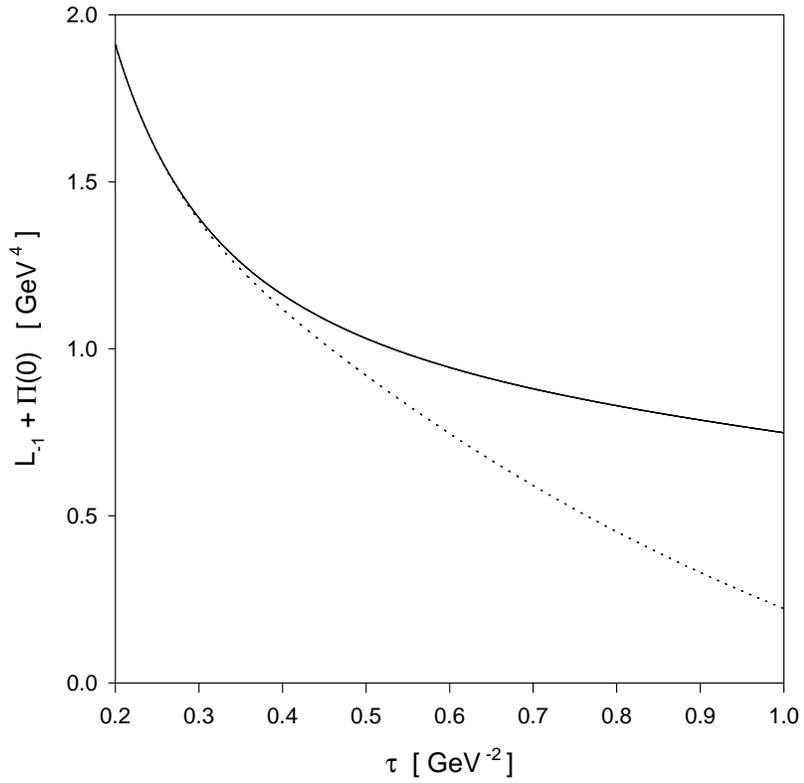}
\caption{
Comparison of the full field-theoretical content of ${\cal L}_{-1}(\tau)+\Pi(0)$ with  (dashed curve) and without
(solid curve) instanton effects.
}
\label{sm1_fig}
\end{figure}

\begin{figure}[htb]
\centering
\includegraphics[scale=0.7]{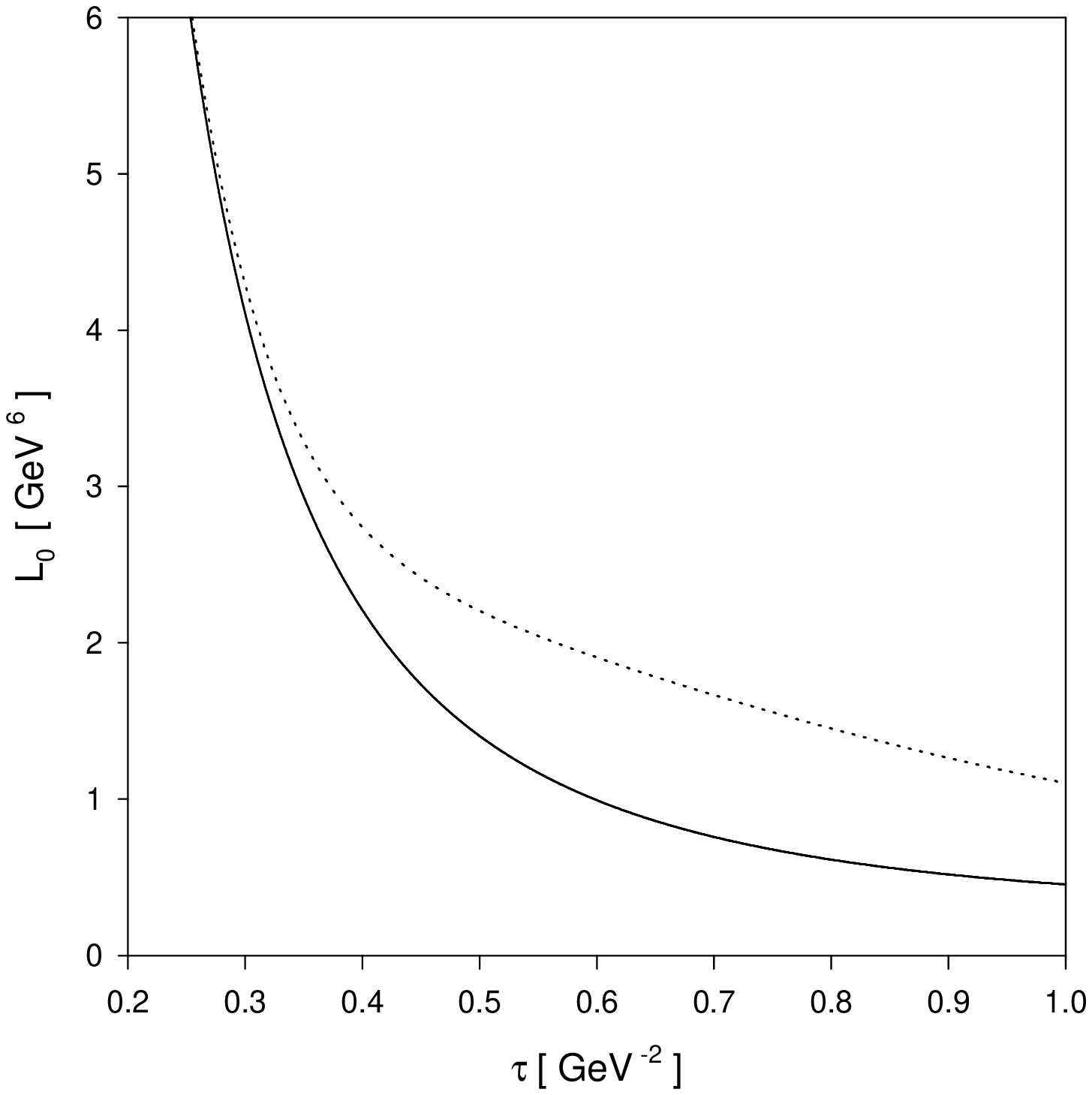}
\caption{
Comparison of ${\cal L}_{0}(\tau)$ with instanton effects included (dashed curve) and instanton effects excluded
(solid curve).
}
\label{s0_fig}
\end{figure}

\begin{figure}[htb]
\centering
\includegraphics[scale=0.7]{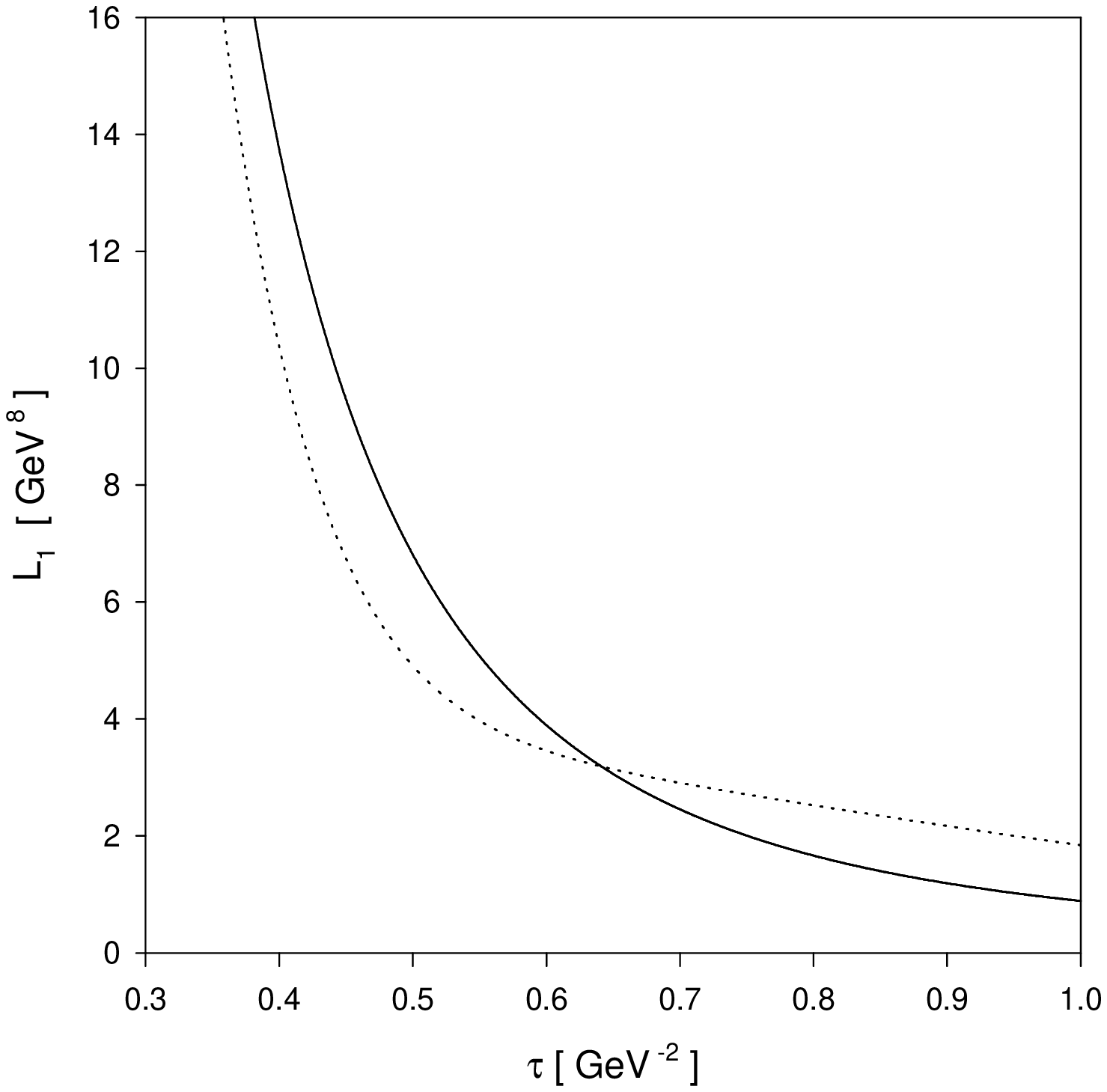}
\caption{
Comparison of ${\cal L}_{1}(\tau)$ with instanton effects included (dashed curve) and instanton effects excluded
(solid curve).
}
\label{s1_fig}
\end{figure}

\begin{figure}[htb]
\centering
\includegraphics[scale=0.7]{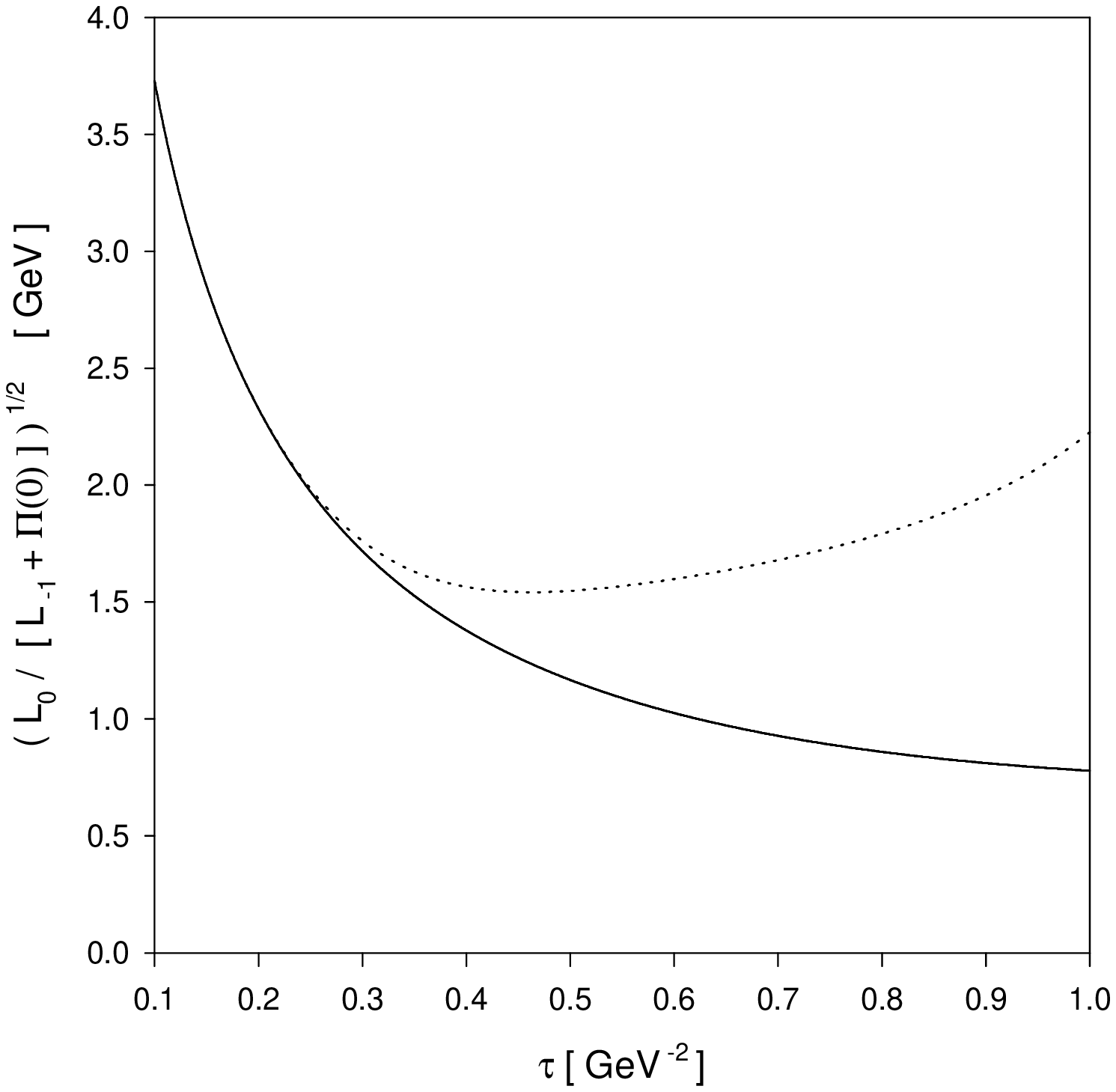}
\caption{
Comparison of the ratio $\sqrt{{\cal L}_{0}(\tau)/\left[{\cal L}_{-1}(\tau)+\Pi(0)\right]}$ with instanton effects 
included (dashed curve) and 
instanton effects excluded (solid curve).
}
\label{s0sm1_fig}
\end{figure}

\begin{figure}[htb]
\centering
\includegraphics[scale=0.7]{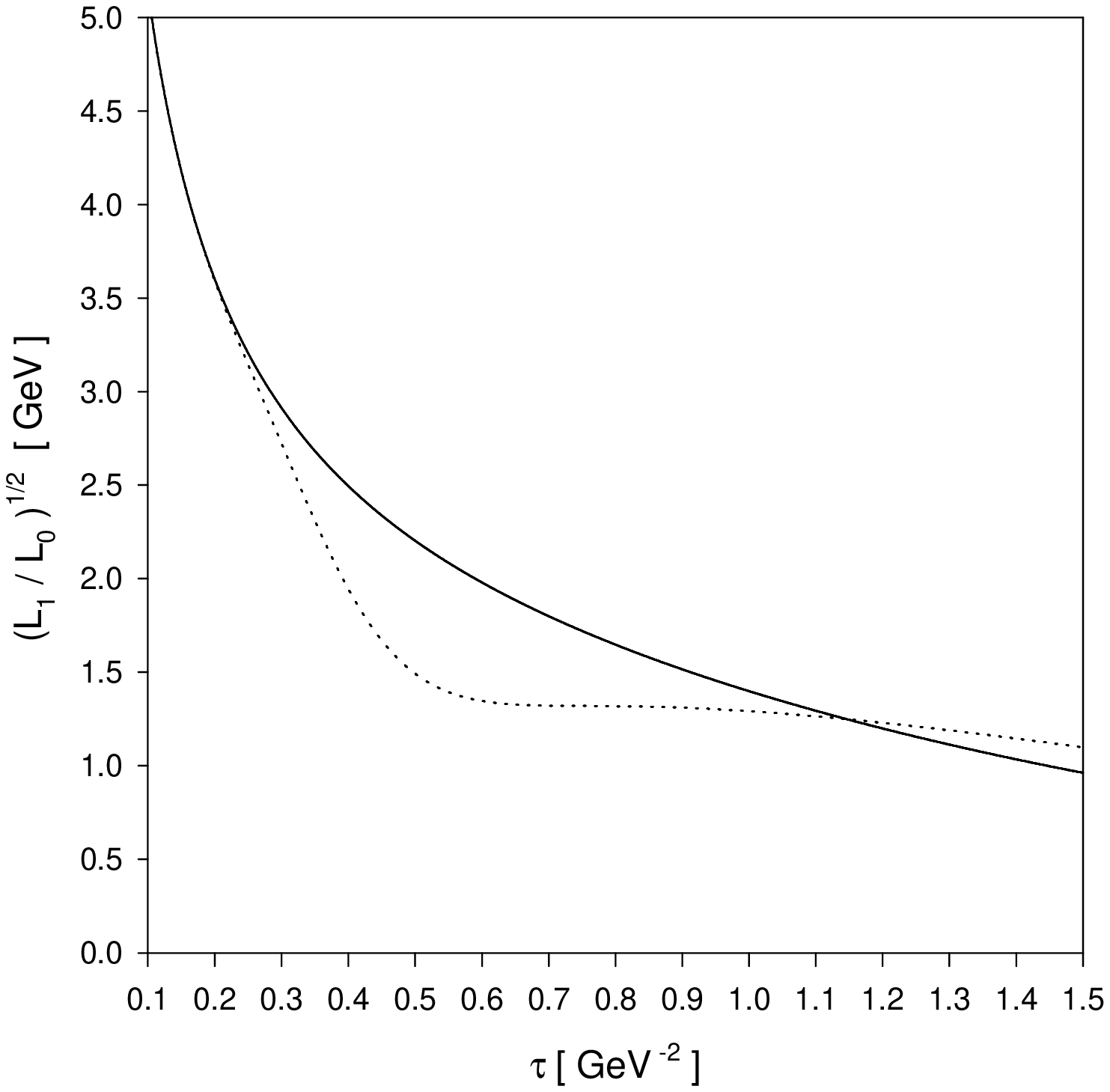}
\caption{
Comparison of the ratio $\sqrt{{\cal L}_{1}(\tau)/{\cal L}_{0}(\tau)}$ with instanton effects included (dashed curve) and 
instanton effects excluded (solid curve).
}
\label{s1s0_fig}
\end{figure}

The nonperturbative QCD parameters are needed for further analysis of the sum-rules.
We employ the DIL model~\cite{ins_liquid} parameters summarized in~\bref{ins_liquid_params},
as well as vacuum saturation  for the dimension-8 gluon condensate~\cite{NSVZ_glue,vacuum_saturation}
\begin{equation}
\left\langle {\cal O}_8\right\rangle=14\left\langle\left(\alpha f_{abc}G^a_{\mu\rho}G^b_{\nu\rho}\right)^2\right\rangle
-\left\langle\left(\alpha f_{abc}G^a_{\mu\nu}G^b_{\rho\lambda}\right)^2\right\rangle=
\frac{9}{16}\left(\left\langle \alpha G^2\right\rangle\right)^2
\label{vac_sat}
\end{equation}
and instanton estimates of the dimension-six condensate \cite{NSVZ,SVZ} 
\begin{equation}
\left\langle {\cal O}_6\right\rangle=
\left\langle g f_{abc}G^a_{\mu\nu}G^b_{\nu\rho}G^c_{\rho\mu}\right\rangle
=\left(0.27\unt{GeV^2}\right)\left\langle \alpha G^2\right\rangle\quad .
\end{equation}
Finally,  again using the approximation $\left\langle J\right\rangle=\left\langle \alpha G^2\right\rangle$ and the
central gluon condensate value [see \bref{GG_cond}] from reference \cite{GG_ref}, we find that
the role of instanton contributions  to the sum-rules (\ref{Lm1}--\ref{L1})
is as illustrated in Figures \ref{sm1_fig}, \ref{s0_fig}, and \ref{s1_fig}.  
In particular, 
we see that the instanton contributions  
diminish ${\cal L}_{-1}(\tau)+\Pi(0)$. 
The  LET term $\Pi(0)$, which leads to the asymptotic flattening of ${\cal L}_{-1}(\tau)+\Pi(0)$
at a value substantially different from zero when instantons are absent, is clearly suppressed by 
 instanton effects in the large $\tau$ region.  As noted earlier, such flattening over the $\tau\le 1.0\,{\rm GeV^{-2}}$
region would be indicative via (\ref{lap_m1_res}) of a sub-GeV lowest-lying resonance ({\it i.e.}, $m_r^2\tau\ll 1$ 
over the physically relevant region of $\tau$)
Instanton effects no only undo this flattening, but also increase ${\cal L}_0(\tau)$ and
alter the shape of ${\cal L}_1(\tau)$. 
The corresponding effects of instantons on the sum-rule ratios (\ref{L1_bounds}--\ref{L0_bounds}) is shown in 
Figures \ref{s0sm1_fig} and \ref{s1s0_fig}.  
As expected from the instanton's impact of lowering  ${\cal L}_{-1}(\tau)$  and elevating
 ${\cal L}_{0}(\tau)$, the ratio  ${\cal L}_{0}(\tau)/\left[{\cal L}_{-1}(\tau)+\Pi(0)\right]$ is increased substantially
by inclusion of instanton effects,  increasing the corresponding upper bound on the mass of the lightest 
gluonium state.   Instanton effects also serve to  {\em lower} the ratio  
${\cal L}_{1}(\tau)/{\cal L}_{0}(\tau)$, decreasing the corresponding upper bound on the mass of the lightest 
gluonium state.

Figures \ref{ratios_on_fig} and \ref{ratios_off_fig} summarize the ratio (mass bound) analysis in the
 presence and in the 
absence of instanton effects.  It is evident that instanton effects lead to a substantial increase in 
the mass bound on the 
lightest gluonium state, but 
other important features emerge.  For example, the instanton suppression of the LET term $\Pi(0)$ 
reduces the discrepancy between the
ratios including or omitting the $k=-1$ sum-rule.  Furthermore, 
a $\tau$-minimum stability plateau crucial for establishing a credible upper mass bound 
is seen to occur at 
reasonable energy scales ($1/\sqrt{\tau}\le 1.0\unt{GeV}$) only when instanton effects are included. 
The ratios with instanton effects included (see Figure \ref{ratios_on_fig}) are remarkably flat, suggesting
 that the mass bounds could be close to the mass prediction that would be obtained from a 
full sum-rule analysis incorporating the QCD continuum ({\it i.e.} $s_0<\infty$) in the phenomenological 
model.

\section{Discussion}
\label{discuss_sec}
We have calculated the instanton contribution to the Laplace sum-rules of scalar gluonium
and demonstrated explicitly how, for the lowest order $k=-1$ sum-rule, this instanton
contribution cancels part of the dominant LET constant.

As noted in the Introduction, a discrepancy between the lowest lying states evident from the
lowest and from the next-to-lowest Laplace sum-rules may be indicative of two distinct states. 
Such is found to be the case, for example, in the pseudoscalar channel in which the pion
dominates the leading ($k = 0$) Laplace sum rule, but the  $\Pi(1300)$ resonance is found to dominate
the next-to-leading ($k = 1$) Laplace sum rule, because of a mass-suppression 
of the lowest-lying (pion) state
in the latter sum rule
 \cite{pi_1300}. Moreover, analyses of the scalar gluonium
channel in the absence of explicit instanton contributions seem to exhibit a similar discrepancy
between leading $k=-1$ and non-leading $k>-1$ sum-rules \cite{dispersion,NSVZ_glue,glue}.

Prior QCD sum-rule analyses of the instanton contribution to the scalar gluonium channel have
focused either on the $k = -1$ sum rule exclusively \cite{shuryak} or the LET-insensitive 
$k>-1$ sum-rules \cite{forkel}.  Although these two analyses (which are separated by almost two decades)
are both indicative of a lowest lying-resonance mass near or above $1.4\unt{GeV}$, their input content
(parameter values and levels of perturbation theory) are necessarily different, suggesting the need
for a single consistent treatment of leading $k=-1$ and non-leading $k>-1$ Laplace sum-rules in the scalar
gluonium channel that is inclusive of instanton effects. 
We have shown here that careful consideration of the contribution arising from instantons within
the $k = -1$ sum rule in this channel leads to consistency with higher sum-rules in the estimation of
lowest-lying resonance masses in the scalar gluonium channel.\footnote{Note also that we have incorporated the
significant NNLO perturbative corrections, as opposed to  LO corrections in \protect\cite{shuryak} and NLO corrections in
\protect\cite{forkel}.}

\begin{figure}[htb]
\centering
\includegraphics[scale=0.7]{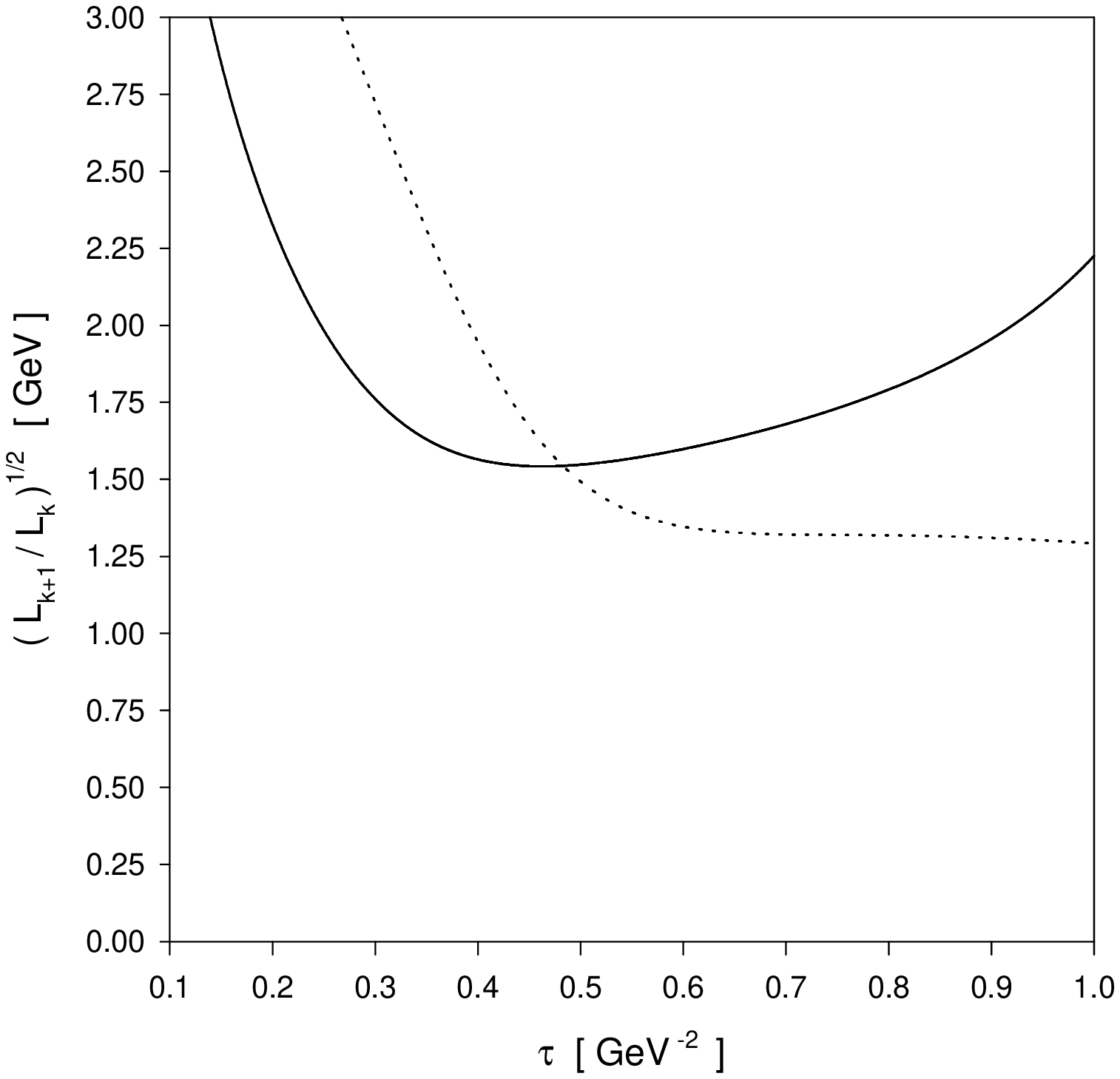}
\caption{
Sum-rule ratios used to obtain scalar gluonium mass bounds with inclusion of instanton effects.
The solid curve represents the ratio  $\sqrt{{\cal L}_{0}(\tau)/\left[{\cal L}_{-1}(\tau)+\Pi(0)\right]}$ 
and the dashed curve represents the ratio  $\sqrt{{\cal L}_{1}(\tau)/{\cal L}_{0}(\tau)}$. 
}
\label{ratios_on_fig}
\end{figure}

\begin{figure}[htb]
\centering
\includegraphics[scale=0.7]{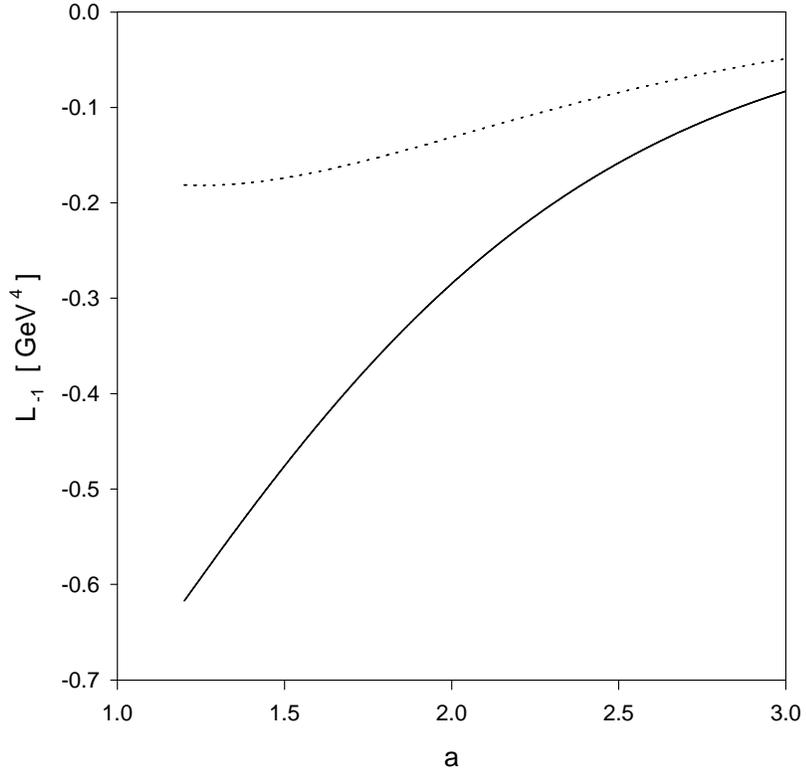}
\caption{
Instanton contributions to ${\cal L}_{-1}$ in the instanton liquid model as a function of $a=\rho_c^2/(2\tau)$.
The solid line represents the complete expression (\protect\ref{full_inst}) for the instanton contributions,
and the dashed line represents the lead term of the large $a$ expansion   (\protect\ref{shuryak_limit}) obtained in 
\protect\cite{shuryak}.
}
\label{shuryak_compare_fig}
\end{figure}

Correspondence with the prior treatment of the $k = -1$ sum rule \cite{shuryak} can be obtained by
examining the instanton contribution \bref{sm1_inst} to this sum rule in the high-energy limit of small $\tau$. 
This contribution is obtained through evaluation of the integral in \bref{sm1_inst}
in the large-$a$ [$\tau  = \rho^2/(2a)$] limit:
\begin{eqnarray}
{\cal L}_{-1}^{inst}(\tau) &=&
-128\pi^2\int dn(\rho)
-32\pi^3\int dn(\rho) \int\limits_0^\infty x^3 J_2\left(x\right)
Y_2\left(x\right)e^{-\frac{x^2}{2a}} \, dx
\nonumber\\
& =& -64\pi^2\int dn(\rho) 
\left[
\frac{\rho^4}{4\tau^2}\left[1+\frac{\rho^2}{2\tau}\right]K_0\left(\frac{\rho^2}{2\tau}\right)+
\frac{\rho^2}{2\tau}\left[2+\frac{\rho^2}{\tau}+\frac{\rho^4}{4\tau^2}\right] K_1\left(\frac{\rho^2}{2\tau}\right)\right]
e^{-\frac{\rho^2}{2\tau}}
\label{full_inst}\\
&\rightarrow & -16\pi^{\frac{5}{2}}\int dn(\rho)  e^{-\frac{\rho^2}{\tau}}\rho^5\tau^{-\frac{5}{2}}
\left[1+{\cal O}\left(\frac{\tau}{\rho^2}\right)\right]~,~\tau\ll\rho^2~.
\label{shuryak_limit}
\end{eqnarray}
In the instanton liquid model [$\,dn(\rho ) = n_c \delta\left(\rho -  \rho_c\right)d\rho\,$],  
\bref{shuryak_limit} is consistent with eq.\ (42) of reference \cite{shuryak} for the instanton contribution 
to the $k = -1$ sum-rule, which was
utilized to anticipate the ${\cal}(1.6 \unt{GeV})$ lattice prediction of the scalar gluonium mass.  
Figure \ref{shuryak_compare_fig}
provides a comparison of this asymptotic form with the exact expression \bref{full_inst} 
for ${\cal L}_{-1}^{inst}(\tau)$              
under  the instanton-liquid assumption. Of
particular interest is the difference between the two expressions over the range $2 \lesssim  a \lesssim  3$,
corresponding to the $0.4\unt{GeV^{-2}}\lesssim\tau\lesssim 0.6\unt{GeV^{-2}}$ range in Figure \ref{ratios_on_fig} 
for which $\sqrt{{\cal L}_0/[\mathcal{L}_{-1}+\Pi(0)]}$ is flat.  Although both
the asymptotic expression of \cite{shuryak} and the exact expression provide negative
contributions which mitigate dominance of the positive LET contribution over ${\cal L}_{-1}$, the 
exact expression is substantially larger in magnitude over the region of phenomenological
interest. We speculate that such an increase relative to the analysis of \cite{shuryak} serves to 
compensate for the larger phenomenological value at present for the gluon condensate within the
low-energy theorem term \bref{PI_approx}, although it may also compensate the non-leading
perturbative contributions in \bref{pi_pert} not known at the time of \cite{shuryak}.

\begin{figure}[htb]
\centering
\includegraphics[scale=0.7]{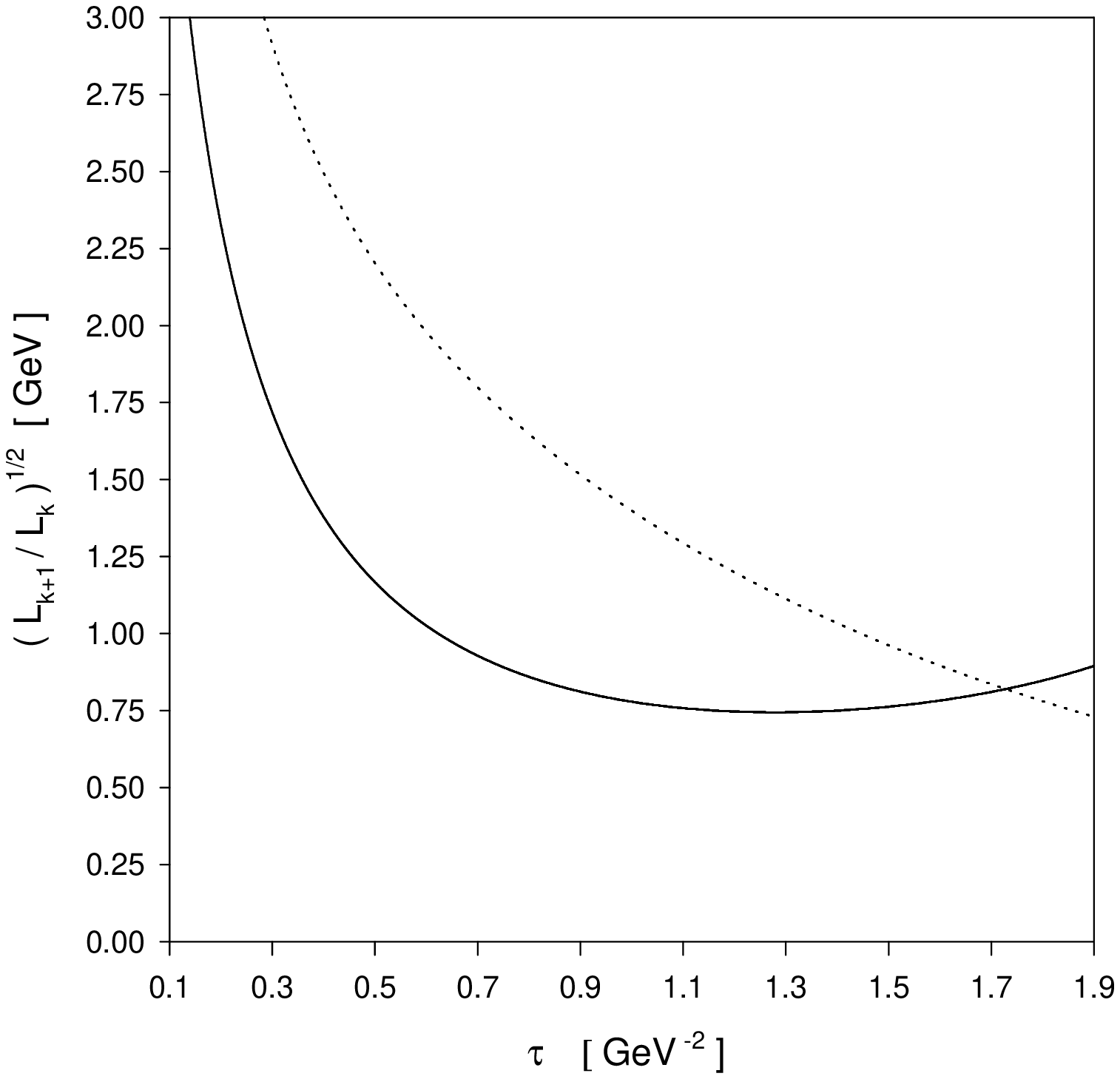}
\caption{
Sum-rule ratios used to obtain scalar gluonium mass bounds with omission of instanton effects.
The solid curve represents the ratio  $\sqrt{{\cal L}_{0}(\tau)/\left[{\cal L}_{-1}(\tau)+\Pi(0)\right]}$ 
and the dashed curve represents the ratio  $\sqrt{{\cal L}_{1}(\tau)/{\cal L}_{0}(\tau)}$. 
}
\label{ratios_off_fig}
\end{figure}

Of course, more sophisticated expressions for the instanton contributions have been utilized to generate
consistent scalar gluonium phenomenology both within \cite{forkel} and complementary to 
\cite{multi_instanton} a QCD sum rule framework, as noted above. The key point of the work presented
here is the reconciliation of the $k = -1$ sum rule with higher-$k$ sum-rules. We reiterate the approximate
consistency between scalar-gluonium masses obtained from the flattened regions of the 
$\sqrt{{\cal L}_0/[\mathcal{L}_{-1}+\Pi(0)]}$
and $\sqrt{{\cal L}_1/{\cal L}_0}$ curves of Figure \ref{ratios_on_fig}, 
as well as the drastic reduction of the scalar gluonium mass
evident in the curves of Figure \ref{ratios_off_fig} when instanton contributions are omitted.  
Roughly speaking, such
contributions account for half the lowest-lying scalar-gluonium mass within a sum-rule context.

\smallskip
\noindent
{\bf Acknowledgements:}  The authors  are grateful for research support
from the Natural Sciences and Engineering Research Council of Canada (NSERC).
We are also grateful to  N. Kochelev for helpful correspondence.

\clearpage

\clearpage

\clearpage

\clearpage

\end{document}